\documentstyle[preprint,aps,epsf,graphicx,psfig]{revtex}
\tighten
\newcommand{\bff}[1]{{\mbox{\boldmath $#1$}}}
\begin{document}
\title{The time-dependent relativistic mean-field theory and
the random phase approximation}
\author{P. Ring$^{1}$,
Zhong-yu Ma$^{2}$\thanks{also Institute of Theoretical Physics,
Beijing, P.R. of China}~,
Nguyen Van Giai$^3$,
D. Vretenar$^{1}$\thanks{On leave from University of Zagreb, Croatia}~,\\
A. Wandelt$^{1}$, and Li-gang Cao$^{2}$
}
\address{1. Physics Department, TU Munich, D-85748 Garching,
Germany \\
2. China Institute of Atomic Energy, Beijing, P.R. of China\\
3. Institut de Physique Nucl\'eaire, IN2P3-CNRS, F-91406 Orsay Cedex, France
\\
}
\date{January 25, 2001}
\maketitle

\begin{abstract}
The Relativistic Random Phase Approximation (RRPA) is derived from
the Time-dependent Relativistic Mean Field (TD RMF) theory in
the limit of small amplitude oscillations.
In the {\it no-sea} approximation of the RMF theory,
the RRPA configuration space includes not only the ususal particle-hole
$ph$-states, but also $\alpha h$-configurations,
i.e. pairs formed from occupied states
in the Fermi sea and empty negative-energy states in the Dirac sea.
The contribution of the negative energy states to the RRPA matrices
is examined in a schematic model, and the large effect of Dirac sea states
on isoscalar strength distributions is illustrated for the
giant monopole resonance in $^{116}$Sn. It is shown that, because
the matrix elements of the time-like component of the vector meson fields
which couple the $\alpha h$-configurations with the
$ph$-configurations are strongly reduced
with respect to the corresponding matrix elements of the isoscalar
scalar meson field, the inclusion of states
with unperturbed energies more than 1.2 GeV below the Fermi energy
has a pronounced effect on giant resonances
with excitation energies in the MeV region. The influence of
nuclear magnetism, i.e. the effect of the
spatial components of the vector fields is examined, and the
difference between the non-relativistic and relativistic
RPA predictions for the nuclear matter compression modulus is explained.
\end{abstract}



\vspace{1cm}

%

\section{Introduction}

Models based on the Relativistic Mean Field (RMF) approximation provide
a microscopic self-consistent description of nuclear structure phenomena
(for reviews, see Refs. \cite{SW.86,Rei.89,Ser.92,Rin.96}). In this
framework classical equations of motion are derived self-consistently from
a fully relativistic Lagrangian. Vacuum polarization effects, as well as
Fock exchange terms, are usually not taken into account explicitly.
This framework, however, is based on an effective theory and the
parameters of effective interactions are determined from a set
of experimental data. In adjusting the parameters of the effective
Lagrangian, a large part of vacuum polarization effects and effects
of exchange terms are already taken into account. In fact, both
contributions have been treated explicitly in nuclear matter and in some
finite spherical nuclei
\cite{HS.83,BMG.87,BFG.93,HS.84,Was.91}, and it has been shown that
these contributions are not small. However, by renormalizing the
parameters of the effective Lagrangian, virtually identical
results for ground-state properties have been obtained without
the inclusion of vacuum polarization and exchange terms.
Essential for a quantitative description of properties of complex
nuclei are the non-linear terms in the meson sector \cite{BB.77}, which
in a simple way include an effective density dependence of the
meson coupling parameters.

The relativistic mean-field models have been mostly applied in the
description of ground-state properties of nuclei all over the periodic table.
In several cases this framework has also been very successfully
applied to excited states. For example, the cranked relativistic mean-field
model \cite{KR.89} describes a large variety of phenomena in
rotational bands of superdeformed nuclei \cite{AKR.96,ALR.98,AKR.99}.
Another example is the Time-dependent RMF model which has been used
to describe the dynamics of giant resonances in
nuclei \cite{VBR.95,PVR.96,VLB.97}. From the time evolution of
multipole moments of the single-particle density, the excitation
energies of giant resonances can be determined. Excellent agreement
with experimental values of isoscalar and isovector giant monopole,
giant quadrupole and the isovector giant dipole resonances has been
obtained. The disadvantage of the time-dependent approach is that
in some cases the large computational effort prevents an accurate
description of excited states as, for instance, the dynamics of
low-lying collective excitations.

The Relativistic Random Phase Approximation (RRPA)
represents the small amplitude limit of the
time-dependent relativistic mean-field theory.
Some of the earliest applications of the RRPA
\cite{Fur.85,HG.89,BC.88,DF.90,HP.90} to finite nuclei include
the description of low-lying negative parity excitations in
$^{16}$O~\cite{Fur.85}, and studies of isoscalar
giant resonances in light and medium nuclei~\cite{HG.89}.
These RRPA calculations, however, were based on the most simple,
linear $\sigma - \omega$ relativistic mean field model.
Only recently non-linear meson self-interaction terms have
been taken included in RRPA calculations \cite{MGT.97,MTG.97}.
However, in these calculations the RRPA configuration space included
only ordinary particle-hole pairs. This seems a reasonable approximation,
since the states formed from occupied positive-energy states in the
Fermi sea and empty negative energy states in the Dirac sea,
have unperturbed energies more than 1.2 GeV below the Fermi level.
It turned out, however, that excitation energies of isoscalar
resonances calculated in this way were very different from those
obtained in the TD RMF approach with the same effective interactions
\cite{GMT.99,VRL.99}.

Since it is well known that in the non-relativistic framework the RPA
corresponds to the small
amplitude limit of time-dependent Hartree-Fock (TDHF) (see, e.g., Ref. \cite
{RS.80}), these discrepancies remained an open puzzle for a couple of years.
Only recently it has been shown \cite{VWR.00,MGW.00} that, in order to
reproduce results of time-dependent relativistic mean-field calculations
for giant resonances, the RRPA configuration space must contain
negative-energy states from the Dirac sea. In principle, if the Dirac sea
were fully occupied, these configurations
would be forbidden by the Pauli principle.
However, in the {\it no-sea} approximation the negative-energy states
do not contribute to the nucleon densities, i.e. these states are
not occupied. It is, thus, possible to form $\alpha h$ pairs ($\alpha$
empty state in the Dirac sea, $h$ occupied state in the Fermi sea)
and include them in the RRPA configuration space. Although formally
possible, and also necessary in order to preserve symmetries,
this procedure raises some serious conceptual problems because the $\alpha h$
configurations have negative unperturbed excitations energies. This means
that the energy surface is no longer positive definite. The static solutions
of the RMF equations correspond to saddle points on the energy surface,
rather than to minima as in non-relativistic Hartree-Fock calculations.
It is not, therefore, a priori clear that small amplitude oscillations
around these stationary solutions will be stable. Furthermore, it is
not obvious why configurations with unperturbed energies more than
1.2 GeV below the Fermi level have such a pronounced effect on the
excitation energies of giant resonances in the MeV region.
In non-relativistic calculations, for instance, it has been noted \cite{RS.74}
that $ph$-configurations with very large excitation energies affect
the position of the spurious mode, but they have no effect on the
excitation energies of giant resonances.

The purpose of the present investigation is to clarify some of these
problems and to understand in a better way the relation between
the RRPA and the TD RMF in the {\it no-sea} approximation.
The paper is organized as follows: the time-dependent RMF model
is analyzed in Sec. II. In Sec. III the RRPA is derived from the
TDRMF equations in the limit of small amplitude motion.
In particular, we discuss the importance of $\alpha h$-pairs in the
RRPA configuration space. In Sec. IV we introduce a relativistic
extension of the Brown-Bolsterli model and solve it in the
linear response approximation. It is shown how large matrix elements,
which couple the $\alpha h$-sector with the $ph$-configurations,
can arise in the RRPA matrices.
The large effect of Dirac sea states on isoscalar strength
distributions is illustrated in Sec. V. The results are summarized
in Sec. VI.


\section{The time-dependent relativistic mean-field model}

In quantum hadrodynamics
the nucleus is described as a system of Dirac nucleons which
interact through the
exchange of virtual mesons and photons. The model is
based on the one-boson exchange description of the
nucleon-nucleon interaction. The Lagrangian
density reads \cite{SW.86}
\begin{eqnarray}
{\cal L} &=&\bar{\psi}\left( i\gamma \cdot \partial -m\right) \psi ~+~\frac{1%
}{2}\partial \sigma \cdot \partial \sigma -\frac{1}{2}m_{\sigma }\sigma ^{2}
\nonumber \\
&&-~\frac{1}{4}\Omega _{\mu \nu }\Omega ^{\mu \nu }+\frac{1}{2}m_{\omega
}^{2}\omega ^{2}~-~\frac{1}{4}{\vec{{\rm R}}}_{\mu \nu }{\vec{{\rm R}}}^{\mu
\nu }+\frac{1}{2}m_{\rho }^{2}\vec{\rho}^{\,2}~-~\frac{1}{4}{\rm F}_{\mu \nu
}{\rm F}^{\mu \nu }  \nonumber \\
&&-~\bar{\psi}[g_{\sigma }\sigma ~+~g_{\omega }\gamma \cdot \omega
~+~g_{\rho }\gamma \cdot \vec{\rho}\vec{\tau} ~+~e\gamma \cdot A \frac{%
(1-\tau_{3})}{2}]\psi~.  \label{E2.1}
\end{eqnarray}
Vectors in isospin space are denoted by arrows, and bold-faced
symbols will indicate vectors in ordinary three-dimensional space;
a dot denotes the scalar product in Minkowski space ($\gamma \cdot
\omega =\gamma^\mu\omega_\mu
=\gamma _{0}\omega _{0}-\bff{\gamma}\bff{\omega}$). The Dirac spinor
$\psi$ denotes the nucleon with mass $m$.
$m_\sigma$, $m_\omega$, and $m_\rho$ are the masses of the
$\sigma$-meson, the $\omega$-meson, and the $\rho$-meson,
and $g_\sigma$, $g_\omega$, and $g_\rho$ are the
corresponding coupling constants for the mesons to the
nucleon, and $e^{2}/\hbar c =1/137.036$.
Eq.(\ref{E2.1}), $\Omega ^{\mu \nu }$, $\vec{R}%
^{\mu \nu }$, and $F^{\mu \nu }$ denote the field tensors of the vector fields $%
\omega $, $\rho $, and of the photon, respectively:
\begin{eqnarray}
\Omega ^{\mu \nu } &=&\partial ^{\mu }\omega ^{\nu }-\partial ^{\nu }\omega
^{\mu }~,  \nonumber \\
\vec{R}^{\mu \nu } &=&\partial ^{\mu }\vec{\rho}^{\,\nu }-\partial ^{\nu }%
\vec{\rho}^{\,\mu }~,  \nonumber \\
F^{\mu \nu } &=&\partial ^{\mu }A^{\nu }-\partial ^{\nu }A^{\mu }~.
\label{E2.2}
\end{eqnarray}
If the bare masses $m$, $m_{\omega }$, and $m_{\rho }$ are used for the
nucleons and the $\omega $ and $\rho $ mesons, there are only four free
model parameters: $m_{\sigma }$, $g_{\sigma }$, $g_{\omega }$
and $g_{\rho }$. Their values can be adjusted to experimental
data of just few spherical nuclei. This simple model, however, is
not flexible enough for a quantitative description of properties of
complex nuclear systems. An effective density
dependence has been introduced \cite{BB.77} by replacing the quadratic
$\sigma $-potential $\frac{1}{2}m^2_{\sigma }\sigma ^{2}$
with a quartic potential $U(\sigma )$
\begin{equation}
U(\sigma )~=~\frac{1}{2}m_{\sigma }^{2}\sigma ^{2}+\frac{1}{3}g_{2}\sigma
^{3}+\frac{1}{4}g_{3}\sigma ^{4}~.  \label{E2.3}
\end{equation}
The potential includes the nonlinear $\sigma $
self-interaction, with two additional  parameters $g_{2}$ and $g_{3}$.
The corresponding Klein-Gordon equation
becomes nonlinear, with a $\sigma $%
-dependent mass $m^2_{\sigma }(\sigma )=$ $m^2_{\sigma }+g_{2}\sigma
+g_{3}\sigma ^{2}$. More details on the relativistic mean-field
formalism can be found in Refs.~\cite{SW.86,Rei.89,Ser.92,Rin.96}.

From the Lagrangian density the set of coupled equations of
motion is derived. The Dirac equation for the nucleons
\begin{eqnarray}
i\partial _{t}\psi _{i} &=&\hat{h}(t)\psi _{i}  \nonumber \\
&=&\left\{\bff{\alpha}\lbrack-i\bff{\nabla}-
{\bf V(r},t{\bf )]+}V({\bf r},t)+
{\bf \beta }\big(m-S({\bf r},t)\big)\right\} \psi _{i}.
\label{E2.4}
\end{eqnarray}
If one neglects retardation effects for the meson fields, the
time-dependent mean-field potentials
\begin{eqnarray}
S({\bf r},t) &=&-g_{\sigma }\sigma ({\bf r},t)~,  \nonumber \\
V_{\mu }({\bf r},t) &=&g_{\omega }\omega _{\mu }({\bf r},t)+g_{\rho }\vec{%
\tau}\vec{\rho}_{\mu }({\bf r},t)+eA_{\mu }({\bf r},t)\frac{(1-\tau _{3})}{2}%
~,  \label{E2.5}
\end{eqnarray}
are calculated at each step in time from the solution of
the stationary Klein-Gordon equations
\begin{eqnarray}
\left[ -\Delta +m_{\sigma }^{2}\right] \,\sigma ({\bf r,}t) &=&-g_{\sigma
}\,\rho _{s}({\bf r,}t)-g_{2}\,\sigma ^{2}({\bf r,}t)-g_{3}\,\sigma ^{3}(%
{\bf r,}t)~,  \nonumber \\
\left[ -\Delta +m_{\omega }^{2}\right] \,\omega _{\mu }({\bf r,}t)
&=&g_{\omega }\,j_{\mu }({\bf r,}t)~,  \nonumber \\
\left[ -\Delta +m_{\rho }^{2}\right] \,\vec{\rho}_{\mu }({\bf r,}t)
&=&g_{\rho }\,\,\vec{j}_{\mu }({\bf r},t)~,  \nonumber \\
-\Delta \,A_{\mu }({\bf r,}t) &=&e\,j_{c\mu }({\bf r},t)~.  \label{E2.6}
\end{eqnarray}
This approximation is justified by the large masses
in the meson propagators. Retardation effects can be neglected
because of the short range of the corresponding meson exchange forces.
In the mean-field approximation only the motion of the nucleons
is quantized, the meson degrees of freedom are described by
classical fields which are defined by the nucleon densities and
currents. The single-particle spinors $\psi_i~(i=1,2,...,A)$
form the A-particle Slater determinant $|\Phi(t)\rangle$.
The nucleons move independently in the classical meson fields,
i.e. residual two-body correlations are not included, and
the many-nucleon wave function is a Slater determinant at
all times. The sources of the fields in the Klein-Gordon
equations are the nucleon densities and currents
calculated in the {\it no-sea} approximation
\begin{eqnarray}
\rho _{s}({\bf r},t) &=&\sum\limits_{i=1}^{A}\bar{\psi}_{i}^{{}}({\bf r}%
,t)\psi _{i}^{{}}({\bf r},t)~,  \nonumber \\
j_{\mu }({\bf r},t) &=&\sum\limits_{i=1}^{A}\bar{\psi}_{i}^{{}}({\bf r}%
,t)\gamma _{\mu }\psi _{i}^{{}}({\bf r},t)~,  \nonumber \\
\vec{j}_{\mu }({\bf r},t) &=&\sum\limits_{i=1}^{A}\bar{\psi}_{i}^{{}}({\bf r}%
,t)\vec{\tau}\gamma _{\mu }\psi _{i}^{{}}({\bf r},t)~,  \nonumber \\
j_{c\mu }({\bf r},t) &=&\sum\limits_{i=1}^{Z}\bar{\psi}_{i}^{{}}({\bf r}%
,t)\gamma _{\mu }\psi _{i}^{{}}({\bf r},t)~.  \label{E2.7}
\end{eqnarray}
where the summation is over all occupied states in the Fermi sea.
In the {\it no-sea} approximation the negative-energy states
do not contribute to the densities and currents, i.e. vacuum
polarization is explicitly neglected. However, as already discussed
in the Introduction, this is an
effective theory with the parameters of the Lagrangian
determined from a set of experimental data. In adjusting
the parameters of the effective Lagrangian, a large part
of vacuum polarization effects is therefore already taken into
account. It should be emphasized that the {\it no-sea} approximation
is essential for practical applications of the relativistic
mean-field model.

The stationary solutions of the relativistic mean-field equations describe
the ground-state of a nucleus. They correspond to stationary
points on the relativistic energy surface.
The Dirac sea, i.e. the negative energy eigenvectors of the
Dirac hamiltonian, is different
for different nuclei. This means that it depends on the specific
solution of the set of non-linear RMF equations. The Dirac
spinors which describe the ground-state of a finite nucleus (positive
energy states) can be expanded, for instance, in terms of vacuum solutions,
which form a complete set of plane wave functions in spinor space.
This set is only complete, however, if in addition to the positive energy
states, it also contains the states with negative energies,
in this case the Dirac sea of the vacuum. Positive energy solutions
of the RMF equations in a finite nucleus automatically contain vacuum
components with negative energy. In the same way, solutions which
describe excited states, as for instance states with different
angular momenta which are solutions of the cranked RMF equations,
contain negative energy components which correspond to the ground-state
solution.

This is also true for the solutions of the time-dependent problem.
Although for the stationary solutions the negative-energy states
do not contribute to the densities in the {\it no-sea} approximation,
their contribution is implicitly included in
the time-dependent calculation. The coupled system of RMF equations
describes the time-evolution of A nucleons in the
effective mean-field potential. Starting from the self-consistent solution
which describes the ground-state of the nucleus, initial
conditions can be defined which correspond, for instance, to
excitations of giant resonances
in experiments with electromagnetic or hadron probes.
For example, the one-body proton and neutron densities can
be initially deformed and/or given some initial velocities.
The resulting mean-field dynamics can be described by the time-evolution
of the collective variables. In coordinate space for example,
these will be the multipole moments of the density distributions.
At each time $t$, the Dirac spinors $\psi _{i}(t)$ can be expanded
in terms of the complete set of
solutions of the stationary Dirac equation $\psi _{k}^{(0)}$
\begin{equation}
\psi _{i}({\bf r},t)=\sum_{k}c_{k}(t)\psi _{k}^{(0)}({\bf r})\text{e}%
^{-i\varepsilon _{k}t}+\sum_{\alpha}c_{\alpha }(t)\psi _{\alpha }^{(0)}(%
{\bf r})\text{e}^{-i\varepsilon _{\alpha }t}~,  \label{E2.9}
\end{equation}
where the index $k$ runs over all positive energy eigen-solutions $\varepsilon _{k}>0$
(hole states $h$ in the Fermi sea, and particle states $p$ above the Fermi sea),
and the index $\alpha $ denotes eigen-solutions with negative energy
$\varepsilon _{\alpha }<0$. We follow the time evolution of $A$
Dirac spinors which at time $t=0$ form the Fermi sea of the stationary solution.
This means that at each time we have a {\it local} Fermi sea of $A$
time-dependent spinors which, of course, contain components of
negative-energy solutions of the stationary Dirac equation. One could also
start with the infinitely many negative energy solutions
$\psi _{\alpha}({\bf r},t=0)$ ($\varepsilon_\alpha <0$), and propagate
them in time with the same hamiltonian $\hat{h}(t)$. Since the
time-evolution operator is unitary \cite{RS.80}
\begin{equation}
i\partial _{t}\left\langle \psi _{i}|\psi _{a}\right\rangle =\left\langle
\psi _{i}|h^{\dagger }-h|\psi _{a}\right\rangle =0,  \label{E2.9a}
\end{equation}
the states which form the {\it local} Dirac sea are orthogonal to the
{\it local} Fermi sea at each time. This is the meaning of the
{\it no-sea} approximation in the time-dependent problem.
For small-amplitude oscillations around the stationary solution,
the coefficients $c_{\alpha }(t)$ of the negative energy components in
(\ref{E2.9}) are, of course, small.

We will first consider linear relativistic mean-field models
($g_{2}=g_{3}=0$ in (\ref{E2.3})). In the instantaneous approximation, i.e.
neglecting the time derivatives $\partial_t^2$ in the Klein-Gordon equations,
the solutions for the mean-fields are calculated from
\begin{eqnarray}
\sigma ({\bf r},t) &=&g_{\sigma }\int D_{\sigma }({\bf r,r}^{\prime })\rho
_{s}({\bf r}^{\prime },t)d^{3}r~,  \nonumber \\
\omega _{\mu }({\bf r},t) &=&g_{\omega }\int D_{\omega }({\bf r,r}^{\prime
})j_{\mu }({\bf r}^{\prime },t)d^{3}r~,  \nonumber \\
\vec{\rho}_{\mu }({\bf r},t) &=&g_{\rho }\int D_{\rho }({\bf r,r}^{\prime })%
\vec{j}_{\mu }({\bf r}^{\prime },t)d^{3}r~,  \nonumber \\
A_{\mu }({\bf r},t) &=&e\int D_{photon}({\bf r,r}^{\prime })j_{c\mu }({\bf r}%
^{\prime },t)d^{3}r~.  \label{E2.10}
\end{eqnarray}
The propagators have the Yukawa form
\begin{equation}
D_{\phi }({\bf r},{\bf r}^{\prime })\,=\pm \frac{1}{4\pi }\frac{\text{e}%
^{-m_{\phi }|{\bf r}-{\bf r}^{\prime }|}}{|{\bf r}-{\bf r}^{\prime }|}~,
\label{E2.11}
\end{equation}
where $\phi $ denotes the mesons $\sigma $, $\omega $, $\rho $,
and the photon. The plus (minus) sign is for vector (scalar) fields.
In the non-linear case an analytic solution of the Klein-Gordon
equation is, of course, no longer possible. The corresponding meson
field is a non-linear functional of the density and currents.

The relativistic single-particle density matrix reads
\begin{equation}
\hat{\rho}({\bf r},{\bf r}^{\prime },t)=\sum\limits_{i=1}^{A}|\psi _{i}^{{}}(%
{\bf r},t)\rangle \langle \psi _{i}^{{}}({\bf r}^{\prime },t)|~.
\label{E2.12}
\end{equation}
If the Dirac spinor is written in terms of large and small components
\begin{equation}
|\psi _{i}^{{}}({\bf r},t)\rangle =\left(
\begin{array}{c}
\,\,\,\,f_{i}({\bf r},t) \\
ig_{i}({\bf r},t)
\end{array}
\right),   \label{E2.13}
\end{equation}
the density matrix takes the form
\begin{equation}
\rho ({\bf r},{\bf r}^{\prime },t)=\left(
\begin{array}{cc}
\,\,\,\sum\limits_{i=1}^{A}f_{i}^{{}}({\bf r},t)f_{i}^{\dagger }({\bf r}%
^{\prime },t) & -i\sum\limits_{i=1}^{A}f_{i}^{{}}({\bf r},t)g_{i}^{\dagger }(%
{\bf r}^{\prime },t) \\
i\sum\limits_{i=1}^{A}g_{i}^{{}}({\bf r},t)f_{i}^{\dagger }({\bf r}^{\prime
},t) & \,\,\,\,\,\,\sum\limits_{i=1}^{A}g_{i}^{{}}({\bf r},t)g_{i}^{\dagger
}({\bf r}^{\prime },t)
\end{array}
\right) ~.  \label{E2.14}
\end{equation}
Further, a relativistic two-body interaction is defined
\begin{equation}
\hat{V}=\int d^{3}r_{1}d^{3}r_{2}\hat{\psi}^{\dagger }({\bf r}_{1})\hat{\psi}%
^{\dagger }({\bf r}_{2})V({\bf r}_{1},{\bf r}_{2})\hat{\psi}({\bf r}_{1})%
\hat{\psi}({\bf r}_{2})~,  \label{E2.15}
\end{equation}
where $\hat{\psi}^{\dagger }$and $\hat{\psi}$ are the Dirac field creation and
annihilation operators, and
\begin{equation}
V({\bf r}_{1},{\bf r}_{2})=D_{\sigma }({\bf r}_{1},{\bf r}_{2})\,\beta
^{(1)}\beta ^{(2)}+D_{\omega }^{{}}({\bf r}_{1},{\bf r}_{2})
\left( 1-\bff{\alpha}^{(1)}\bff{\alpha}^{(2)}\right) ~.
\label{E2.16}
\end{equation}
In order to simplify the notation, we omit the $\rho $-meson and the photon,
though they are, of course, included in actual applications of the
relativistic mean-field model. Their contribution to the matrix elements
of $V({\bf r}_{1},{\bf r}_{2})$ is, however, much smaller than that of
the $\sigma$ and $\omega$ mesons.

By introducing an arbitrary complete spinor basis (the indices $k,l,\dots $
denote both positive and negative energy states), the two-body interaction
operator can be written in the form
\begin{equation}
\hat{V}=\frac{1}{2}\sum_{kk^{\prime }ll^{\prime }}V_{klk^{\prime }l^{\prime
}}\hat{\psi} _{k}^{\dagger }\hat{\psi} _{l}^{\dagger }
\hat{\psi} _{l^{\prime }}^{{}}\hat{\psi}
_{k^{\prime }}^{{}}~.  \label{E2.17}
\end{equation}
The single-particle equation of motion corresponds to the time-dependent
relativistic Hartree problem
\begin{equation}
i\partial _{t}\psi _{i}=\hat{h}(\hat{\rho})\psi _{i}~,  \label{E2.18}
\end{equation}
with the Dirac Hamiltonian
\begin{equation}
\hat{h}(\hat{\rho})=
\bff{\alpha}\,{\bf p} + \beta(m+\,\Sigma(\hat{\rho})),
\label{E2.19}
\end{equation}
and the mass operator
\begin{equation}
\Sigma _{kl}(\hat{\rho})=\sum_{k^{\prime }l^{\prime }}V_{kl^{\prime
}lk^{\prime }}\rho _{k^{\prime }l^{\prime }}~.  \label{E2.20}
\end{equation}
The corresponding equation of motion for the density operator reads
\begin{equation}
i\partial _{t}\hat{\rho}=\left[ \hat{h}(\hat{\rho}),\hat{\rho}\right] ~,
\label{E2.21}
\end{equation}
in full analogy with the non-relativistic Hartree-Fock problem (see, e.g.,
Ref. \cite{RS.80}).

In expressing the TD RMF equations (\ref{E2.4}-\ref{E2.6}) in terms of a
relativistic two-body interaction, we have eliminated the meson degrees of
freedom by using the Yukawa form (\ref{E2.11}) of the meson propagators. This
applies, of course, only to Lagrangians
that do not contain non-linear meson self-interactions.
The non-linear couplings are, however, essential for a realistic description
of nuclear properties. Formally, also in this case the Klein-Gordon
equations can be solved at each step in time, and the resulting
meson fields are non-linear functionals of the densities and currents.
The Dirac operator has still the form of Eq. (\ref{E2.19}), but the
mass-operator $\Sigma _{kl}(\hat{\rho})$ becomes a much more complicated
functional of the single-particle density.

The numerical solution of the full time-dependent problem
with non-linear meson self-interactions does not present particular
difficulties (see Refs. \cite{VBR.95,PVR.96,VLB.97}). Much more
difficult, however, is to
eliminate the meson degrees of freedom and to derive a relativistic two-body
interaction in the general case of large amplitude motion. This has only been
done in the small amplitude limit \cite{MGT.97}. The $\sigma $-field
and the scalar density $\,\rho _{s}$ are expanded in the neighborhood
of the stationary ground-state solutions
\begin{eqnarray}
\sigma ({\bf r,}t{\bf )} &=&{\bf \,}\sigma ^{(0)}({\bf r)+}\delta \sigma (%
{\bf r,}t{\bf )}~,  \label{E2.22} \\
\,\rho _{s}({\bf r,}t{\bf )} &=&{\bf \,}\rho _{s}^{(0)}({\bf r)+}\delta \rho
_{s}({\bf r,}t{\bf )}~.
\end{eqnarray}
The corresponding Klein-Gordon equation (\ref{E2.6}) for the $\sigma $-field
is solved by linearization, i.e. up to terms linear in $\delta \sigma$ we obtain
\begin{equation}
\left[ -\Delta +m_{\sigma }^{2}({\bf r})\right] \delta \,\sigma ({\bf r,}%
t)=-g_{\sigma }\delta \,\rho _{s}({\bf r,}t)~,  \label{E2.23}
\end{equation}
with
\begin{equation}
m_{\sigma }^{2}({\bf r})=\left. \frac{\partial ^{2}U}{\partial \sigma ^{2}}%
\right| _{\sigma =\sigma ^{(0)}({\bf r})}~.  \label{E2.24}
\end{equation}
The $\sigma $-meson propagator is defined by the equation
\begin{equation}
\left[ -\Delta +m_{\sigma }^{2}({\bf r})\right] D_{\sigma }({\bf r},{\bf r}%
^{\prime })=-\delta ({\bf r-r}^{\prime })~,  \label{E2.25}
\end{equation}
and it has been determined numerically in the RRPA calculations of Refs. (
\cite{MGT.97,MTG.97,GMT.99}). In the following only the small amplitude
limit will be studied, and therefore we do not need to worry about
the more general problem of large amplitude motion.


\section{The small amplitude limit of TD RMF and the relativistic RPA}

In this section we study the response of the density matrix $\hat{\rho}(t)$
to an external one-body field
\begin{equation}
\hat{F}(t)=\hat{F}\text{e}^{-i\omega t}+h.c.~,  \label{E3.1}
\end{equation}
which oscillates with a small amplitude. Assuming that in the single-particle
space this field is represented by the operator
\begin{equation}
\hat{f}(t) = \sum_{kl} \, f_{kl}(t) \; \hat{a}^{\dagger}_k \hat{a}^{}_l ,
\end{equation}
the equation of motion for the density operator is
\begin{equation}
i\partial _{t}\hat{\rho}=\left[ \hat{h}(\hat{\rho})+\hat{f}(t),\hat{\rho}%
\right] ~,  \label{E3.2}
\end{equation}
In the linear approximation the density matrix is expanded
\begin{equation}
\hat{\rho}(t)=\hat{\rho}^{(0)}+\delta \hat{\rho}(t)~,  \label{E3.3}
\end{equation}
where $\hat{\rho}^{(0)}$ is the stationary ground-state density.
From the definition of the density matrix (\ref{E2.12}), it follows
that $\hat{\rho}\left( t\right) $ is a projector at all times, i.e.
$\hat{\rho}\left( t\right) ^{2}=$ $\hat{\rho}\left( t\right) $.
In particular, this means that the eigenvalues of $\hat{\rho}^{(0)}$
are 0 and 1. In the non-relativistic case particle states
above the Fermi level correspond to the eigenvalue 0, and hole states
in the Fermi sea correspond to the eigenvalue 1. In the relativistic
case, one also has to take into account states from the Dirac sea.
In the {\it no-sea} approximation these states are not occupied,
i.e. they correspond to the eigenvalue 0 of the density matrix.
We will work in the basis which diagonalizes $\hat{\rho}^{(0)}$
\begin{equation}
\rho _{kl}^{(0)}=\delta _{kl}\rho _{k}^{(0)}=\left\{
\begin{array}{ll}
0 & \text{for unoccupied states above the Fermi level (index }p\text{)}
\\
1 & \text{for occupied states in the Fermi sea (index }h\text{)\quad } \\
0 & \text{for unoccupied states in the Dirac sea (index }\alpha \text{)}
\end{array}
\right.   \label{E3.4}
\end{equation}
Since $\hat{\rho}(t)$ is a projector at all times, in linear order
\begin{equation}
\hat{\rho}^{(0)}\delta \hat{\rho}+\delta \hat{\rho}\hat{\rho}^{(0)}=\delta
\hat{\rho}~.  \label{E3.5}
\end{equation}
This means that the non-vanishing matrix elements of $\delta \hat{\rho}$ are:
$\delta \rho _{ph}$, $\delta \rho_{hp}$,
$\delta \rho _{\alpha h}$, and $\delta \rho _{h\alpha }$. These are
determined by the solution of the TD RMF equation (\ref{E3.2}). In the linear
approximation the equation of motion reduces to
\begin{equation}
i\partial _{t}\delta \hat{\rho}=\left[ \hat{h}^{(0)},\delta \hat{\rho}\right]
+\left[ \frac{\partial \hat{h}}{\partial \rho }\delta \rho ,\hat{\rho}^{(0)}%
\right] +\left[ \hat{f},\hat{\rho}^{(0)}\right] ~,  \label{E3.6}
\end{equation}
where
\begin{equation}
\frac{\partial \hat{h}}{\partial \rho }\delta \rho =\sum_{ph}\frac{\partial
\hat{h}}{\partial \rho _{ph}}\delta \rho _{ph}+\frac{\partial \hat{h}}{%
\partial \rho _{hp}}\delta \rho _{hp}+\sum_{\alpha h}\frac{\partial \hat{h}}{%
\partial \rho _{\alpha h}}\delta \rho _{\alpha h}+\frac{\partial \hat{h}}{%
\partial \rho _{h\alpha }}\delta \rho _{h\alpha }~.  \label{E3.7}
\end{equation}
In the small amplitude limit $\delta \rho$ will, of course, also display
a harmonic time dependence e$^{-i\omega t}$. Taking into account the fact
that $\hat{h}_{kl}^{(0)}=\delta _{kl}\epsilon
_{k}$ is diagonal in the stationary basis, we obtain
\begin{eqnarray}
(\omega -\epsilon _{p}+\epsilon _{h})\delta \rho _{ph}
&=&f_{ph}+\sum_{p^{\prime }h^{\prime }}V_{ph^{\prime }hp^{\prime }}\delta
\rho _{p^{\prime }h^{\prime }}+V_{pp^{\prime }hh^{\prime }}\delta \rho
_{h^{\prime }p^{\prime }}+\sum_{\alpha^{\prime }h^{\prime }}V_{ph^{\prime
}h\alpha^{\prime }}\delta \rho _{\alpha^{\prime }h^{\prime }}+
V_{p\alpha^{\prime }hh^{\prime }}
\delta \rho_{h^{\prime }\alpha^{\prime }}
\nonumber
\\
(\omega -\epsilon_{\alpha }+\epsilon _{h})\delta \rho _{\alpha h}
&=&f_{\alpha h}+\sum_{p^{\prime }h^{\prime }}V_{\alpha h^{\prime }hp^{\prime
}}\delta \rho _{p^{\prime }h^{\prime }}+V_{\alpha p^{\prime }hh^{\prime
}}\delta \rho _{h^{\prime }p^{\prime }}+\sum_{\alpha ^{\prime }h^{\prime
}}V_{\alpha h^{\prime }h\alpha ^{\prime }}\delta \rho _{\alpha ^{\prime
}h^{\prime }}+V_{\alpha \alpha ^{\prime }hh^{\prime }}\delta \rho
_{h^{\prime }\alpha ^{\prime }}  \nonumber \\
(\omega -\epsilon _{h}+\epsilon _{p})\delta \rho _{hp}
&=&f_{hp}+\sum_{p^{\prime }h^{\prime }}V_{hh^{\prime }pp^{\prime }}\delta
\rho _{p^{\prime }h^{\prime }}+V_{hp^{\prime }ph^{\prime }}\delta \rho
_{h^{\prime }p^{\prime }}+\sum_{\alpha ^{\prime }h^{\prime }}V_{hh^{\prime
}p\alpha ^{\prime }}\delta \rho _{\alpha ^{\prime }h^{\prime }}+V_{h\alpha
^{\prime }ph^{\prime }}\delta \rho _{h^{\prime }\alpha ^{\prime }}
\nonumber
\\
(\omega -\epsilon _{h}+\epsilon_{\alpha })\delta \rho _{h\alpha }
&=&f_{h\alpha }+\sum_{p^{\prime }h^{\prime }}V_{hh^{\prime }
\alpha p^{\prime}}\delta \rho _{p^{\prime }h^{\prime }}+
V_{hp^{\prime }\alpha h^{\prime}}
\delta \rho _{h^{\prime }p^{\prime }}+
\sum_{\alpha ^{\prime }h^{\prime}}
V_{hh^{\prime }\alpha \alpha ^{\prime }}
\delta \rho _{\alpha ^{\prime}h^{\prime }}+
V_{h\alpha ^{\prime }\alpha h^{\prime }}
\delta \rho_{h^{\prime }\alpha ^{\prime }}
\label{E3.8}
\end{eqnarray}
or, in matrix form
\begin{equation}
\left[ \omega \left(
\begin{array}{cc}
1 & 0 \\
0 & -1
\end{array}
\right) -\left(
\begin{array}{cc}
A & B \\
B^{\ast } & A^{\ast }
\end{array}
\right) \right] \left(
\begin{array}{c}
X \\
Y
\end{array}
\right) =\left(
\begin{array}{c}
F \\
\bar{F}
\end{array}
\right) ~,  \label{E3.9}
\end{equation}
The RRPA matrices $A$ and $B$ read
\begin{eqnarray}
A &=&\left(
\begin{array}{cc}
(\epsilon _{p}-\epsilon _{h})\delta _{pp^{\prime }}\delta _{hh^{\prime }} &
\\
& (\epsilon _{\alpha }-\epsilon _{h})\delta _{\alpha \alpha ^{\prime
}}\delta _{hh^{\prime }}
\end{array}
\right) +\left(
\begin{array}{cc}
V_{ph^{\prime }hp^{\prime }} & V_{ph^{\prime }h\alpha ^{\prime }} \\
V_{\alpha h^{\prime }hp^{\prime }} & V_{\alpha h^{\prime }h\alpha ^{\prime }}
\end{array}
\right)   \label{E3.10} \\
B &=&\left(
\begin{array}{cc}
V_{pp^{\prime }hh^{\prime }} & V_{p\alpha ^{\prime }hh^{\prime }} \\
V_{\alpha p^{\prime }hh^{\prime }} & V_{\alpha \alpha ^{\prime }hh^{\prime }}
\end{array}
\right)   \label{E3.11}
\end{eqnarray}
and the amplitudes $X$ and $Y$ are defined
\begin{equation}
X=\left(
\begin{array}{c}
\delta \rho _{ph} \\
\delta \rho _{\alpha h}
\end{array}
\right) ,\quad Y=\left(
\begin{array}{c}
\delta \rho _{hp} \\
\delta \rho _{h\alpha }
\end{array}
\right) ~.  \label{E3.12}
\end{equation}
The vectors which represent the external field contain
the matrix elements
\begin{equation}
F=\left(
\begin{array}{c}
f_{ph} \\
f_{\alpha h}
\end{array}
\right) ,\quad \bar{F}=\left(
\begin{array}{c}
f_{hp} \\
f_{h\alpha }
\end{array}
\right) ~.  \label{E3.13}
\end{equation}
In conventional linear response theory (see, e.g.,
Ref. \cite{RS.80}) the polarization function $\Pi _{pqp^{\prime }q^{\prime
}}(\omega )$ is defined by the response of the
density matrix to an external field with a harmonic time dependence
\begin{equation}
\delta \rho _{pq}=\sum_{p^{\prime }q^{\prime }}\Pi _{pqp^{\prime }q^{\prime
}}(\omega )\,f_{p^{\prime }q^{\prime }}~.  \label{E3.14}
\end{equation}
Its spectral representation reads
\begin{equation}
\Pi _{pqp^{\prime }q^{\prime }}(\omega )=\sum_{\mu }\frac{\langle 0|\psi
_{q^{{}}}^{\dagger }\psi _{p^{{}}}^{{}}|\mu \rangle \langle \mu |\psi
_{p^{\prime }}^{\dagger }\psi _{q^{\prime }}^{{}}|0\rangle }{\omega -E_{\mu
}+E_{0}+i\eta }-\frac{\langle 0|\psi _{p^{\prime }}^{\dagger }\psi
_{q^{\prime }}^{{}}|\mu \rangle \langle \mu |\psi _{q^{{}}}^{\dagger }\psi
_{p^{{}}}^{{}}|0\rangle }{\omega +E_{\mu }-E_{0}+i\eta }~,  \label{E3.15}
\end{equation}
where the index $\mu $ runs over all excited states
\mbox{$\vert$}%
$\mu \rangle $ with energy $E_{\mu }$. In the RPA approximation the polarization
function is obtained by inverting the matrix
\begin{equation}
\Pi (\omega )=\left[ \left(
\begin{array}{cc}
\omega +i\eta  & 0 \\
0 & -\omega -i\eta
\end{array}
\right) -\left(
\begin{array}{cc}
A & B \\
B^{\ast } & A^{\ast }
\end{array}
\right) \right] ^{-1}~.  \label{E3.16}
\end{equation}
$\Pi (\omega )$ is the solution of the linearized Bethe-Salpeter equation
\begin{equation}
\Pi (\omega )=\Pi ^{0}(\omega )+\Pi ^{0}(\omega )V\,\Pi (\omega )~,
\label{E3.17}
\end{equation}
where the free polarization function is given by
\begin{equation}
\Pi _{klk^{\prime }l^{\prime }}^{0}(\omega )=\frac{\rho _{l}^{(0)}-\rho
_{k}^{(0)}}{\omega -\epsilon _{k}+\epsilon _{l}+i\eta }\delta _{kk^{\prime
}}\delta _{ll^{\prime }}~.  \label{E3.18}
\end{equation}
The eigenmodes of the system are determined by the RPA equation
\begin{equation}
\left(
\begin{array}{cc}
A & B \\
-B^{\ast } & -A^{\ast }
\end{array}
\right) \left(
\begin{array}{c}
X \\
Y
\end{array}
\right) _{\mu }=\left(
\begin{array}{c}
X \\
Y
\end{array}
\right) _{\mu }\Omega _{\mu }~.  \label{E3.19}
\end{equation}
In principle, this is a non-Hermitian eigenvalue problem. In the
non relativistic case, however, it can be reduced to a
Hermitian problem of half dimension, if the RPA matrices are real and if $(A+B)$
is positive definite. In this case one can also show that the eigenvalues $%
\Omega _{\mu }^{2}$ are positive, i.e., the RPA eigenfrequencies $\Omega
_{\mu }$ are real (see \cite{RS.80}).

The relativistic case is much more complicated. From Eq. (\ref{E3.10})
we notice that the matrix $(A+B)$ is not positive definite. The
$\alpha h$ configurations have large negative diagonal matrix
elements  $\epsilon _{\alpha h}=\epsilon _{\alpha}-\epsilon _{h}\le -1.2$
GeV, and the RRPA equation can
no longer be reduced to a Hermitian problem of half dimension. In this case
it is also not clear whether the eigenfrequencies are
necessarily real, because the stability matrix
\begin{equation}
{\cal S}=\left(
\begin{array}{cc}
A & B \\
B^{\ast } & A^{\ast }
\end{array}
\right)  \label{E3.20}
\end{equation}
is no longer positive definite. Rather than minima, the
solutions of the RMF equations are
saddle points\cite{Providencia} in the multi-dimensional
energy surface, and the Thouless theorem \cite{Th.61}, which states that a
positive definite stability matrix ${\cal \ S}$ leads to a stable RPA
equation with real frequencies, does not apply.

However, the opposite is not true: if the stability matrix
is not positive definite, it does not automatically follow that the eigenvalues
of the corresponding RPA matrix are not real. In fact, cases like this occur also
in the non relativistic RPA in the neighborhood of phase transitions, where the
interaction $V$ is very large and attractive. The positive energies $%
\varepsilon_p - \varepsilon_h$ on the diagonal of the stability
matrix are not large enough, as compared to the matrix elements of $V$,
to guarantee positive eigenvalues of ${\cal S}$.
In the relativistic case the energies on the diagonal
$\varepsilon_\alpha - \varepsilon_ h$ are negative. Even for small matrix
elements of V the stability matrix ${\cal S}$ will have negative eigenvalues.
However, as long as the diagonal part dominates, i.e. as long as
we are not in a neighborhood of a phase transition, the RRPA
eigenfrequencies are real. This can be easily demonstrated if
instead of the RPA amplitudes $X$ and $Y$, we define the generalized coordinates
$Q$ and momenta $P$
\begin{equation}
Q=\frac{1}{\sqrt{2}}(X-Y^{\ast }),\;\;\;\;\;\;\;P=\frac{i}{\sqrt{2}}%
(X+Y^{\ast })~.  \label{E3.21}
\end{equation}
In the small amplitude limit the time-dependent mean field equations
take the form of classical Hamiltonian equations (for details see
Ref. \cite{RS.80}, Chapt. 12) for the Hamiltonian function
\begin{equation}
{\cal H}(P,Q)=\frac{1}{2}\left(
\begin{array}{cc}
P^{\ast } & -P
\end{array}
\right) {\cal M}^{-1}\left(
\begin{array}{c}
P \\
-P^{\ast }
\end{array}
\right) +\frac{1}{2}\left(
\begin{array}{cc}
Q^{\ast } & Q
\end{array}
\right) {\cal S}\left(
\begin{array}{c}
Q \\
Q^{\ast }
\end{array}
\right),  \label{E3.22}
\end{equation}
with the inertia tensor
\begin{equation}
{\cal M}=\left(
\begin{array}{cc}
A & -B \\
-B^{\ast } & A^{\ast }
\end{array}
\right) ^{-1}~.  \label{E3.23}
\end{equation}
The large negative diagonal matrix elements are also
present in the inertia tensor. If the off-diagonal matrix
elements are not too large, a negative inertia and
a negative curvature will again result in real frequencies.
In all applications of RRPA we have found real frequencies,
though in none of these cases the stability matrix ${\cal S}$
was positive definite. This also explain why
the time-dependent RMF equations have stable solutions which
describe oscillations with real frequencies around the static solution,
although the static solution itself corresponds to a saddle point.

The solution of the RPA equations in configuration space is much more
complicated in the relativistic case. Firstly, because in addition to
the usual $ph$-states, the configuration space includes a large number
of $\alpha h$-states. A further complication arises because
the full non-Hermitian RPA matrix has to be diagonalized,
even in cases when the matrix elements are real.
The usual method \cite{RS.80}, which reduces the dimension of the RPA
equations by half does not apply.

Summarizing the results of this section, we have shown that the
relativistic RPA represents the small amplitude limit of the
time-dependent RMF theory. However, because the RMF theory
is based on the {\it no-sea} approximation, the RRPA configuration
space includes not only the ususal $ph$-states, but also
$\alpha h$-configurations, i.e. pairs formed from occupied states
in the Fermi sea and empty negative-energy states in the Dirac sea.
At each time $t\neq 0$ the occupied positive energy states
can have non-vanishing overlap with both positive and negative
energy solutions calculated at $t=0$. If the density matrix $\hat{\rho}(t)$
is represented in the basis which
diagonalizes the static solution $\hat{\rho}^{(0)}$, it contains not only
the usual components $\delta \hat{\rho}_{ph}$ with a particle
above the Fermi level and a hole in the Fermi sea, but also components $%
\delta \hat{\rho}_{\alpha h}$ with a particle in the Dirac sea and a hole in
the Fermi sea.

One of the important advantages of using the time-dependent variational
approach is that it conserves symmetries. It is well known
from non-relativistic time-dependent mean field theory that symmetries
are connected with zero energy solutions of the RPA, i.e. the Goldstone modes,
and it is one of the advantages of RPA that it restores the symmetries
broken by the mean field. This has already been realized in the early
studies of symmetry conservation in RRPA, and it has been
emphasized by Dawson and Furnstahl in Ref. \cite{DF.90}, that it is essential
to include the $\alpha h$ configuration space in order to bring the
Goldstone modes to zero energy.

However, it was not anticipated that negative energy states in the RRPA
configuration space could have dramatic effects on the excitation
energies of giant resonances, as we will show in the following sections.
It is not obvious that basis states with unperturbed energies more than 1.2
GeV below the Fermi energy, can have a big influence on giant resonances
with excitation energies in the MeV region. In the following section
we will study a simple model which provides a deeper insight into this problem.

\section{A separable model}

The model studied in this section represents a relativistic extension of the
Brown and Bolsterli model\cite{BB.59}, which has played an essential role
in the understanding of the microscopic picture of collective
excitations.

The single-particle basis consists of 4 states, each of them
$\Omega $-fold degenerate ($\nu =1, \ldots \Omega $). The first two states
(1 and 2) correspond to particle levels with the free mass $m$, the states
3 and 4 correspond to the negative energy levels with free mass $-m$.
The model Hamiltonian reads
\begin{equation}
H=H_{0}-\lambda _{s}^{{}}S^{\dagger }S+\lambda _{v}^{{}}V^{\dagger }V~,
\label{E4.1}
\end{equation}
where $H_0$ is the Hamiltonian which describes
free Dirac particles
\begin{equation}
H_{0}=\sum_{i=1}^{A}
\Big(\bf{\alpha}{\bf p}_{i }+ \beta m_{i }+ \frac{1}{2} \sigma
\varepsilon_{i }\Big)~,
\label{E4.2}
\end{equation}
with
\begin{equation}
\alpha =\left(
\begin{array}{cccc}
0 & 0 & 1 & 0 \\
0 & 0 & 0 & 1 \\
1 & 0 & 0 & 0 \\
0 & 1 & 0 & 0
\end{array}
\right) ,\quad \beta =\left(
\begin{array}{cccc}
1 & 0 & 0 & 0 \\
0 & 1 & 0 & 0 \\
0 & 0 & -1 & 0 \\
0 & 0 & 0 & -1
\end{array}
\right) ,\quad \sigma =\left(
\begin{array}{cccc}
1 & 0 & 0 & 0 \\
0 & -1 & 0 & 0 \\
0 & 0 & 1 & 0 \\
0 & 0 & 0 & -1
\end{array}
\right)  \label{E4.3}
\end{equation}
In Eq.(\ref{E4.2}) $m_i = m$ is the free mass of particle $i$, $p_i = p$ denotes its
momentum, and $\varepsilon_{i} = \varepsilon_{0} \ll m$ induces
a small splitting between the levels 1 and 2 (and, of course, between 3 and 4).

The interaction consists of an attractive scalar field $S$ and a repulsive
vector field $V$
\begin{equation}
S=\sum_{i=1}^{A}\left(
\begin{array}{cccc}
0 & 1 & 0 & 0 \\
1 & 0 & 0 & 0 \\
0 & 0 & 0 & -1 \\
0 & 0 & -1 & 0
\end{array}
\right) _{i },\quad \quad V=\sum_{i=1}^{A}\left(
\begin{array}{cccc}
0 & 1 & 0 & 0 \\
1 & 0 & 0 & 0 \\
0 & 0 & 0 & 1 \\
0 & 0 & 1 & 0
\end{array}
\right) _{i}~,  \label{E4.4}
\end{equation}
with the strength parameters $\lambda _{s}$ and $\lambda _{v}.$ In the
formalism of second
quantization the operators $H_{0\text{,}}$ $S$ and $V$ take the forms
\begin{eqnarray}
H_{0} &=&p\sum_{\nu }c_{1\nu }^{\dagger }c_{3\nu }^{{}}+c_{2\nu }^{\dagger
}c_{4\nu }^{{}}+h.c.  \nonumber \\
&& + m\sum_{\nu }c_{1\nu }^{\dagger }c_{1\nu }^{{}}+c_{2\nu }^{\dagger
}c_{2\nu }^{{}}-c_{3\nu }^{\dagger }c_{3\nu }^{{}}-c_{4\nu }^{\dagger
}c_{4\nu }^{{}}  \nonumber \\
&& + \frac{\varepsilon _{0}}{2}\sum_{\nu }c_{1\nu }^{\dagger }c_{1\nu
}^{{}}-c_{2\nu }^{\dagger }c_{2\nu }^{{}}+c_{3\nu }^{\dagger }c_{3\nu
}^{{}}-c_{4\nu }^{\dagger }c_{4\nu }^{{}}~,  \label{E4.5} \\
S &=&\sum_{\nu }c_{1\nu }^{\dagger }c_{2\nu }^{{}}-c_{3\nu }^{\dagger
}c_{4\nu }^{{}}+h.c.~,  \label{E4.6} \\
V &=&\sum_{\nu }c_{1\nu }^{\dagger }c_{2\nu }^{{}}+c_{3\nu }^{\dagger
}c_{4\nu }^{{}}+h.c.  \label{E4.7}
\end{eqnarray}

At the mean field level the diagonalization of the Dirac operator
\begin{equation}
H_{0}=\left(
\begin{array}{cccc}
m+\frac{\varepsilon _{0}}{2} & 0 & p & 0 \\
0 & m-\frac{\varepsilon _{0}}{2} & 0 & p \\
p & 0 & -m+\frac{\varepsilon _{0}}{2} & 0 \\
0 & p & 0 & -m-\frac{\varepsilon _{0}}{2}
\end{array}
\right) ~,  \label{E4.10}
\end{equation}
result in the eigenvalues
\begin{equation}
\epsilon _{p}=E+\frac{\varepsilon _{0}}{2},\quad \quad \epsilon _{h}=E-\frac{%
\varepsilon _{0}}{2},\quad \quad \epsilon _{\alpha }=-E+\frac{\varepsilon
_{0}}{2},\quad \quad \epsilon _{\alpha ^{\prime }}=-E-\frac{\varepsilon _{0}%
}{2},  \label{E4.12}
\end{equation}
and the corresponding eigenvectors are
\begin{equation}
\psi _{p}=\left(
\begin{array}{c}
f \\
0 \\
g \\
0
\end{array}
\right) ,\,\,\,\,\psi _{h}=\left(
\begin{array}{c}
0 \\
f \\
0 \\
g
\end{array}
\right) ,\,\,\,\,\,\psi _{\alpha }=\left(
\begin{array}{c}
-g \\
0 \\
f \\
0
\end{array}
\right) ,\,\,\,\,\,\,\psi _{\alpha ^{\prime }}=\left(
\begin{array}{c}
0 \\
-g \\
0 \\
f
\end{array}
\right) ~,  \label{E4.13}
\end{equation}
respectively. We use the notation
\begin{equation}
E=\sqrt{p^{2}+m^{2},},\,\quad \quad \quad f=\cos \frac{\phi }{2},\,\quad
\quad \quad g=\sin \frac{\phi }{2}~,  \label{E4.14}
\end{equation}
with 
\begin{equation}
\tan \frac{\phi }{2}=\frac{p}{m+E}~.  \label{E4.16}
\end{equation}
A realistic choice of single particle energies is
\begin{eqnarray}
\epsilon _{ph} &=&\epsilon _{p}-\epsilon _{h}=\varepsilon _{0}\approx 10%
\text{ MeV}~,  \nonumber \\
\epsilon _{\alpha h} &=&\epsilon _{\alpha }-\epsilon _{h}=-2E+\varepsilon
_{0}\simeq -2\text{ GeV}~,  \nonumber \\
\epsilon _{\alpha ^{\prime }h} &=&\epsilon _{\alpha ^{\prime }}-\epsilon
_{h}=-2E\simeq -2\text{ GeV}~.  \label{E4.18}
\end{eqnarray}
In realistic calculations the ratio between large and small components
of the Dirac spinors is approximately $f/g\approx 30$, i.e., $\phi \simeq 33^{0}$.
In the basis (\ref{E4.13}) the matrices of the operators $S$ and $V$ are
\begin{equation}
S=\sum_{i}\left(
\begin{array}{cccc}
0 & \cos \phi  & 0 & -\sin \phi  \\
\cos \phi  & 0 & -\sin \phi  & 0 \\
0 & -\sin \phi  & 0 & -\cos \phi  \\
-\sin \phi  & 0 & -\cos \phi  & 0
\end{array}
\right) _{i}~,  \label{E4.19}
\end{equation}
\begin{equation}
V=\sum_{i}\left(
\begin{array}{cccc}
0 & 1 & 0 & 0 \\
1 & 0 & 0 & 0 \\
0 & 0 & 0 & 1 \\
0 & 0 & 1 & 0
\end{array}
\right) _{i}~.  \label{E4.20}
\end{equation}
We notice the essential matrix elements
\begin{equation}
\begin{array}{ll}
S_{ph}=\cos \phi  & V_{ph}=1 \\
S_{\alpha h}=-\sin \phi  & V_{\alpha h}=0 \\
S_{\alpha ^{\prime }h}=0 & V_{\alpha ^{\prime }h}=0~.
\end{array}
\label{E4.21}
\end{equation}
In analogy to the Brown-Bolsterli model, the unperturbed polarization
function is
\begin{equation}
\Pi _{FF^{\prime }}^{0}(\omega )=\frac{2\varepsilon _{0}^{{}}\Omega
F_{ph}^{{}}F_{ph}^{\prime }}{\omega _{{}}^{2}-\varepsilon _{0}^{2}+i\eta }+%
\frac{2\varepsilon _{\alpha h}^{{}}\Omega F_{\alpha h}^{{}}F_{\alpha
h}^{\prime }}{\omega _{{}}^{2}-\varepsilon _{\alpha h}^{2}+i\eta }~,
\label{E4.22}
\end{equation}
where the operators $F,$ $F^{\prime }$ $\in$ \{$S,\ V$\}. In particular,
\begin{eqnarray}
\Pi _{SS}^{0}(\omega ) &=&\frac{2\varepsilon _{0}^{{}}\Omega \cos ^{2}\phi }{%
\omega _{{}}^{2}-\varepsilon _{0}^{2}+i\eta }-\frac{2E\Omega \sin ^{2}\phi }{%
\omega _{{}}^{2}-E_{{}}^{2}+i\eta }~,  \label{E4.23} \\
\Pi _{VV}^{0}(\omega ) &=&\frac{2\varepsilon _{0}^{{}}\Omega }{\omega
_{{}}^{2}-\varepsilon _{0}^{2}+i\eta }~,  \label{E4.24} \\
\Pi _{VS}^{0}(\omega ) &=&\Pi _{VS}^{0}(\omega )=\frac{2\varepsilon
_{0}^{{}}\Omega \cos \phi }{\omega _{{}}^{2}-\varepsilon _{0}^{2}+i\eta }~.
\label{E4.25}
\end{eqnarray}
The RRPA frequencies are the roots of the determinant
\begin{equation}
\det \left( 1-
\begin{array}{cc}
-\Pi _{SS}^{0}(\omega )\lambda _{s} & \Pi _{SV}^{0}(\omega )\lambda _{v} \\
-\Pi _{SV}^{0}(\omega )\lambda _{s} & \Pi _{VV}^{0}(\omega )\lambda _{v}
\end{array}
\right) =0  \label{E4.26}
\end{equation}
The essential difference with respect to
the non-relativistic Brown-Bolsterli model is
the additional term $2E\Omega \sin ^{2}\phi /(\omega ^{2}-E^{2}+i\eta )$ in
the scalar polarization $\Pi _{SS}^{0}(\omega )$.
Without this term (i.e. $\phi =0)$, the eigenfrequencies would be determined by
\begin{equation}
\omega _{{}}^{2}=\varepsilon _{0}^{2}-2\varepsilon _{0}^{{}}\Omega (\lambda
_{s}-\lambda _{v})~,  \label{E4.28}
\end{equation}
with the usual cancellation of scalar and vector interactions. With the
additional term, states with unperturbed energies at $\simeq -2E$ are
included in the RPA configuration space. The interaction between
these states and the $ph$-states is determined by the matrix elements
of the scalar interaction
\begin{equation}
v_{\alpha h^{\prime }hp}=-\lambda _{s}\cos \phi \sin \phi ~.  \label{E4.29}
\end{equation}
These matrix elements are not reduced
by a similar term coming from the vector interaction. In our simplified
model this vector-induced term vanishes as a
consequence of the relativistic structure of the equations:
while for a state from the Fermi sea the large component is the
upper component of the spinor, a state from the Dirac sea
has a large lower component of the spinor.
Due to the  $\gamma$-matrix structure of the vertex,
the matrix elements of the vector interaction vanish.
In realistic calculations these matrix elements
do not vanish identically,
but they are reduced by an order of magnitude as compared to the
corresponding scalar terms.

At excitation energies in the MeV region, $\omega \ll E$ in the denominator
of the second term of Eq.(\ref{E4.23}), and we obtain an energy-independent term
$(2\Omega \lambda_{s}\sin ^{2}\phi )/E$ in the dispersion relation.
The eigenfrequencies are now determined by
\begin{equation}
\omega _{{}}^{2}=\varepsilon _{0}^{2}-\varepsilon _{0}^{{}}2\Omega (\lambda
_{s}-\lambda _{v})+\varepsilon _{0}^{{}}2\Omega \lambda _{s}\sin ^{2}\phi
\frac{E+2\Omega \lambda _{v}}{E+2\Omega \lambda _{s}\sin ^{2}\phi }~,
\label{E4.30}
\end{equation}
i.e. we find an additional repulsion for the collective state.

\section{An illustrative case: the giant monopole resonance}

The RPA equations can be solved either by diagonalizing the RPA
matrix in configuration space (see Eq.(\ref{E3.19})), or the
response function can be calculated by solving
the Bethe-Salpeter equation (\ref{E3.17})
in momentum space\cite{HG.89,MGT.97}. In both cases, of course,
one first has to determine the single-nucleon spinors and
the mean-fields which correspond to the stationary solution for
the ground-state. The Dirac-Hartree
equations and the equations for the meson fields are solved
self-consistently in the mean-field approximation. The eigenvalue
problem is solved, for instance, by diagonalization in a spherically
symmetric harmonic oscillator basis~\cite{GRT.90}. From the spectrum
of single-nucleon states the RPA configuration space is built:
particle-hole ($ph$) and antiparticle-hole ($\alpha h$) pairs
which obey the selection rules for angular momentum, parity and isospin.
The number of basis states is also determined by two cut-off
parameters: the maximal $ph$-energy ($\epsilon _{m}-\epsilon
_{i}<E_{\max }$) and the minimal $\alpha h$-energy ($\epsilon _{\alpha}-\epsilon
_{i}>E_{\min }$). With this basis the RPA matrix is calculated
for the same effective interaction that determines the ground-state,
or the free polarization function $\Pi ^{(0)}$ is calculated in the
response function method. Both methods require that the
single-particle continuum is discretized. In order to smooth out
the RPA strength function, the discrete strength distribution
is folded by a Lorentzian of width $\Gamma $. In the response function method
the folding is automatic if a finite value parameter $i\Gamma $
is used in the denominators of Eqs.(\ref{E3.15}-\ref{E3.18}),
instead of the infinitesimal parameter $i\eta $. We have verified
that identical results are obtained with both methods.

The large effect of Dirac sea states on isoscalar strength
distributions is illustrated in Fig.~1, where we display
the isoscalar monopole RRPA strength in $^{116}$Sn calculated with
the NL3 effective interaction~\cite{LKR.97} and the
width of the Lorentzian is $\Gamma $~=~2~MeV. Recent experimental
data are available for the isoscalar giant monopole resonance
in $^{116}$Sn~\cite{YCL.99}. The solid curve represents the full
RRPA strength and it displays a pronounced peak at 16 MeV,
in excellent agreement with the measured value of 15.9 MeV\cite{YCL.99}.
Giant monopole resonances in spherical nuclei are in best
agreement with experimental data, when calculated with effective Lagrangians
with a nuclear matter compression modulus in the range 250-270 MeV
\cite{VLB.97,VWR.00,MGW.00}. The nuclear matter incompressibility of the
NL3 effective interaction is 272 MeV.

The long-dashed curve in Fig.~1 corresponds to the to the case with no
${\alpha}h$ pairs in the RRPA configuration space.
We notice that, without the contribution
from Dirac sea states, the strength distribution is shifted to
lower energy. The position of the peak is shifted from
$\approx 16$ MeV to below 10 MeV if ${\alpha}h$ pairs
are not included in the RRPA basis. Quantitatively similar
results are also obtained with other effective interactions.
In Fig.~1 we have also separated the contributions of vector and scalar
mesons to the ${\alpha}h$ matrix elements. The dash-dot-dot (dash-dot)
curve corresponds to calculations in which
only vector mesons (scalar mesons) were included in the coupling
between the Fermi sea and Dirac sea states. Both interactions
were included in the positive energy particle-hole matrix elements.
The resulting strength distributions nicely illustrate the
dominant contribution of the isoscalar scalar sigma meson
to the ${\alpha}h$ matrix elements, in complete agreement with
the result obtained in the previous section for
the schematic Brown-Bolsterli model.

It is also interesting to examine the effect of the coupling via the
spatial components of the vector meson fields,
i.e. the term $-\bff{\alpha}^{(1)}\bff{\alpha}^{(2)}$ in the interaction
of Eq. (\ref{E2.16}). In
time-dependent calculations this coupling results from the
nucleon currents. In Fig.2 we display the isoscalar
monopole RRPA strength in $^{116}$Sn calculated as follows:
a) full RRPA (solid curve); b) without the matrix elements
of the spatial components of the vector meson fields (dot-dashed curve);
c) without the contribution of the Dirac sea to the
matrix elements of the spatial components of the vector meson fields
(dashed curve); and d) the free Hartree response function.
The currents do not contribute to the static polarizability and to the
$M_{-1}$ moment. At finite frequencies, however, their contribution
is attractive and it lowers the ISGMR energy by $\approx 2$ MeV..
Therefore, if the contribution of the spatial components of the vector
fields is neglected, a better agreement with experimental values
would be obtained with a lower nuclear matter incompressibility:
$K_{\infty }\simeq 230$ MeV. Incidentally, this lower value
for the nuclear matter incompressibility is the one advocated
by non-relativistic RPA calculations~\cite{BBD.95,CGB.00}.
It has already been noted in time-dependent RMF calculations~\cite{VLB.97},
as well as in recent relativistic RPA studies~\cite{MGW.00}, that
effective interactions which reproduce the IS GMR excitation energies
in finite nuclei have a somewhat higher nuclear matter incompressibility
than the corresponding non-relativistic Skyrme or Gogny interactions.
Here we point to a possible solution to this puzzle:
the current terms in the matrix elements of the
particle-hole interaction (\ref{E3.10},\ref{E3.11}) are given by
\begin{equation}
\langle p||j_{1}(kr)[\bff{\alpha}Y_{1}({\bf \hat{r}})]_{J=0}||h\rangle
=\langle p||j_{1}(kr)\left(
\begin{array}{cc}
0 & [\bff{\sigma}Y_{1}]_{J=0} \\
\lbrack \bff{\sigma}Y_{1}]_{J=0} &
\end{array}
\right) ||h\rangle~,   \label{E5.1}
\end{equation}
which is a typical relativistic term because it couples large and small
components of a Dirac spinor. Since they change parity, terms of the type $[{\bf %
\sigma }Y_{1}]_{J=0}$ cannot contribute to the giant monopole resonance
in a non-relativistic calculation.


\section{Conclusions}

In the last couple of years, several discrepancies have been reported
between the results obtained with the Relativistic Random Phase approximation
and the Time-Dependent Relativistic Mean Field theory, when applied
to the the description of small amplitude collective motion in
atomic nuclei.

In order to resolve this puzzle, in the present work we have
derived the RRPA from the TDRMF equations in the limit of
small amplitude motion. The relativistic single particle density
matrix $\hat{\rho}(t)$ has been expanded in terms of the
stationary solutions of the ground-state. We have shown that
the {\it no-sea} approximation, which is essential for practical application
of the RMF theory in finite nuclei, leads to a fundamental difference
between the relativistic and non-relativistic approaches.
While in the non-relativistic case the time-dependent variation
of the density $\delta \hat{\rho}(t)=\hat{\rho}(t)-\hat{\rho}^{(0)}$
has only $ph$-matrix elements (particle ($p$) above the Fermi surface,
hole ($h$) in the Fermi sea), in the relativistic case
$\delta \hat{\rho}$ contains also $\alpha h$-matrix elements,
where $\alpha$ denotes unoccupied states in the Dirac sea.
The fact that states in the Dirac sea can be occupied is a direct
consequence of the {\it no-sea} approximation. In constructing
the matrix $\delta \hat{\rho}$ one has to take into account that
a complete basis of single particle states contains both
positive and negative energy solutions of the Dirac equation.
Already in Ref.~\cite{DF.90} it has been
shown that an RRPA calculation, consistent with the mean-field
model in the $no-sea$ approximation, necessitates configuration
spaces that include both particle-hole pairs and pairs formed
from occupied states and negative-energy states. The contributions
from configurations built from occupied positive-energy states and
negative-energy states are essential for current conservation and
the decoupling of the spurious state.

What is less obvious, however, is that the inclusion of negative-energy
single particle states in the RRPA configuration space has such a
dramatic effect on the calculated excitation energies of isoscalar
giant resonances. In a schematic model we have shown that, due to
the relativistic structure of the RPA equations, the matrix elements
of the time-like component of the vector meson fields
which couple the $\alpha h$-configurations with the
$ph$-configurations vanish. In realistic calculations these
matrix elements do not vanish identically, but they are strongly reduced
with respect to the corresponding matrix elements of the isoscalar
scalar meson field. As a result, the well known cancellation between
the contributions of the  $\sigma $ and $\omega $ fields, which,
for instance, leads to ground-state solution, does not take place
and we find large matrix elements coupling the $\alpha h$-sector
with the $ph$-configurations. In addition,
the number of $\alpha h$-configurations which can couple to the
$ph$-configurations in the neighborhood of the Fermi surface
is much larger than the number of $ph$-configurations. This can
increase the effect by enhancing the collectivity of the
$\alpha h$-configuration space.

The large effect of Dirac sea states on isoscalar strength
distributions has been illustrated for the giant monopole
resonance in $^{116}$Sn. We have also shown that currents
cannot be neglected in the calculation of giant resonances.
Of course they do not occur in the static case, i.e.
in the calculations of the static polarizability or the $M_{-1}$
moment. At finite frequencies, however, time reversal invariance
is broken and spatial components of the vector meson fields
play an important role. This effect is known as {\it nuclear magnetism}.
It is a genuine relativistic effect, because the matrix elements
couple the large and small components of a Dirac spinor.
Since the spatial components of the vector fields have the
form $-\bff{\alpha}^{(1)}\bff{\alpha}^{(2)}$
$D_{\omega }({\bf r}_{1},{\bf r}_{2})$, they result
in an attractive contribution which lowers the value of the
calculated excitation energies of giant resonances. This explains
the difference between the non-relativistic and relativistic RPA
results for the isoscalar giant monopole resonances in spherical
nuclei, and the corresponding predictions for the nuclear matter
compression modulus.

\bigskip

{\bf ACKNOWLEDGMENTS}

P.R. acknowledges the support and the hospitality extended to him
during his stay at the IPN-Orsay, where a large part of this work was
completed. This work has been supported in part by the
Bundesministerium f\"{u}r Bildung und Forschung under the project 06 TM 979
and by the Deutsche Forschungsgemeinschaft.
It was also partially supported by the National Natural Science Foundation
of China under grant No. 19847002, 19835010-10075080 and Major State Basic
Research Development Program under contract No. G200077407.

\begin{figure}[tbp]
\includegraphics*[scale=0.5, angle=0.]{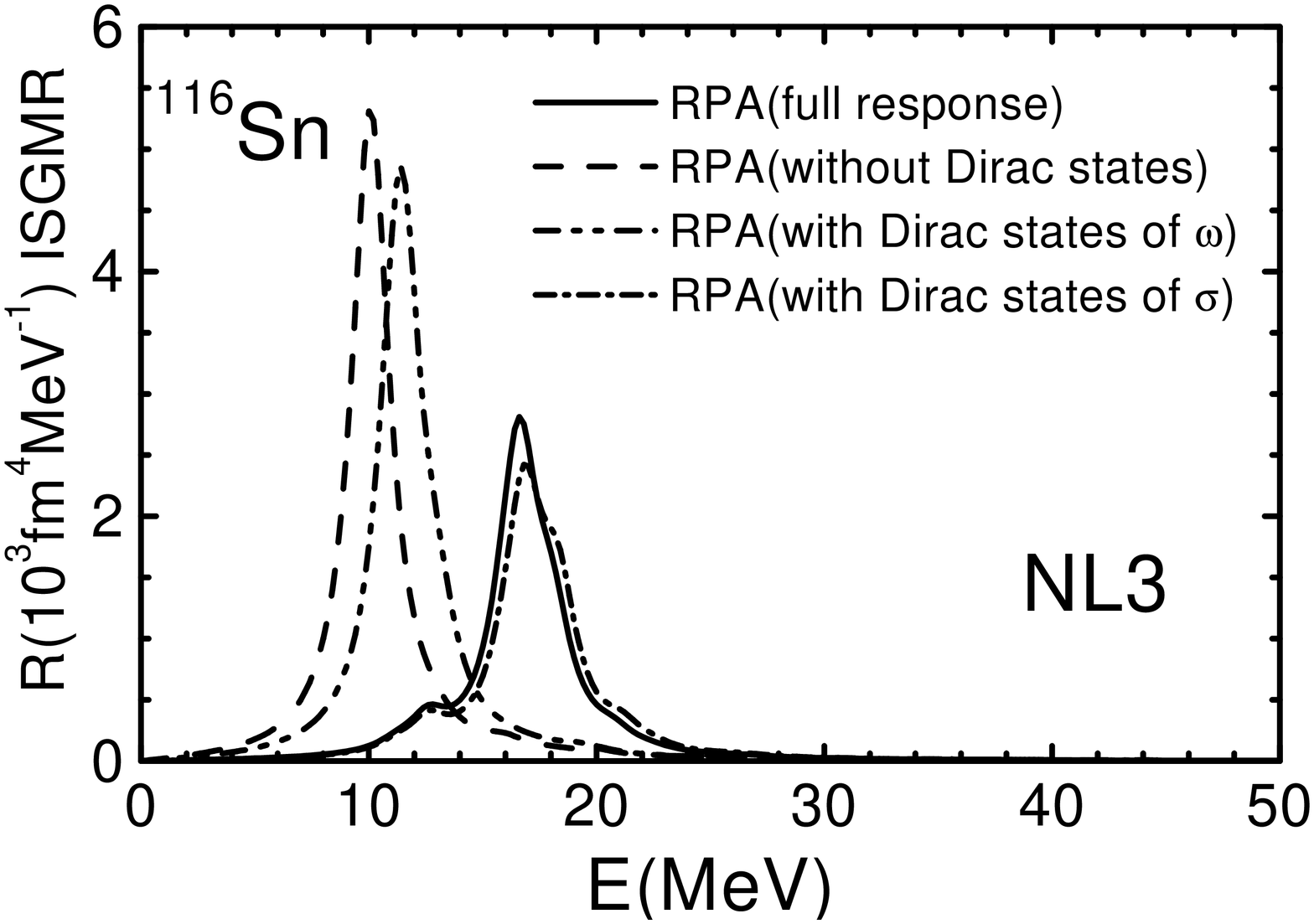}
\caption{ISGMR strength distributions in $^{116}$Sn calculated with
the NL3 effective interaction.
The solid and long-dashed curves are the RRPA
strengths with and without the inclusion of Dirac sea states, respectively.
The dash-dot-dot (dash-dot) curve corresponds to calculations in which
only vector mesons (scalar mesons) are included in the coupling
between the Fermi sea and Dirac sea states.}
\label{Fig.1}
\end{figure}

\begin{figure}[tbp]
\includegraphics*[scale=0.5, angle=0.]{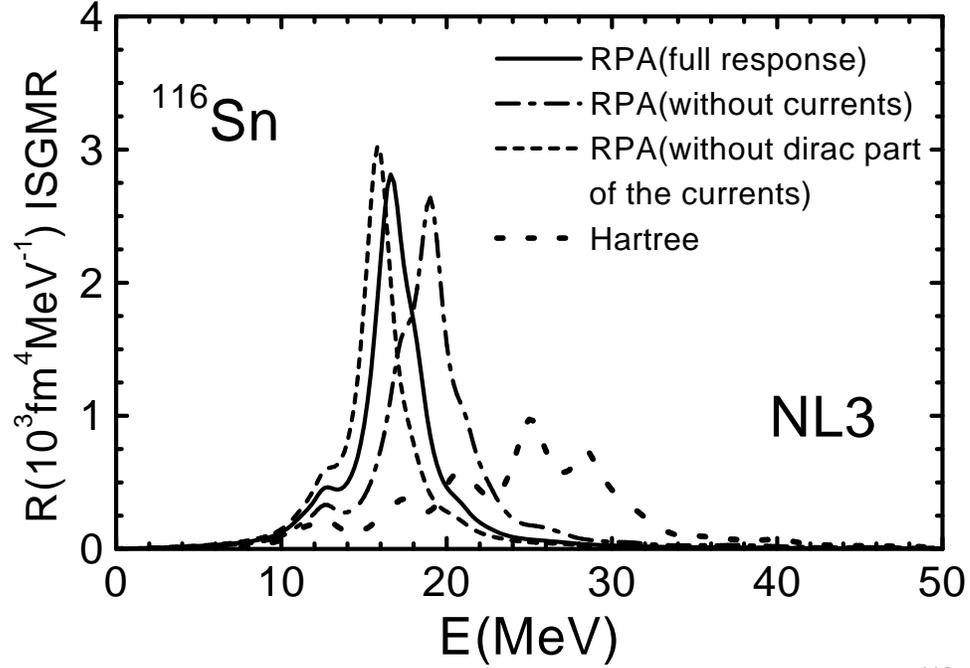}
\caption{Effects of nuclear magnetism on the IS GMR strength distribution in $%
^{116}$Sn. Solid curve: full RRPA calculation; dash-dotted curve: without
the matrix elements
of the spatial components of the vector meson fields; dashed curve:
without the contribution of the Dirac sea to the
matrix elements of the spatial components of the vector meson fields.
The free Hartree response is also displayed.}
\label{Fig.2}
\end{figure}

\end{document}